\documentclass[reprint,amsmath,amssymb,prb,showpacs,superscriptaddress]{revtex4-1}

\usepackage{graphicx}
\usepackage{dcolumn}
\usepackage{bm}

\begin{document}

\title{Structural Modification and Metamagnetic Anomaly in the Ordered State of CeOs$_2$Al$_{10}$}

\author{Yuji Muro}
 \email{ymuro@hiroshima-u.ac.jp}
\author{Jumpei Kajino}%
\affiliation{Department of Quantum matter, AdSM, Hiroshima University, Higashi-Hiroshima, 739-8530, Higashi-Hiroshima, 739-8526, Japan}%

\author{Kazunori Umeo}
\affiliation{Cryogenics and Instrumental Analysis Division, N-BARD, Hiroshima University, Higashi-Hiroshima, 739-8526, Japan}%

\author{Kazue Nishimoto}
\author{Ryuji Tamura}
\affiliation{Department of Materials Science and Technology, Tokyo University of Science, Noda, 278-8510, Japan}%

\author{Toshiro Takabatake}
\affiliation{Department of Quantum matter, AdSM, Hiroshima University, Higashi-Hiroshima, 739-8530, Higashi-Hiroshima, 739-8526, Japan}%
\affiliation{Institute for Advanced Materials Research, Hiroshima University, Higashi-Hiroshima, 739-8530, Japan}%

\date{\today}

\begin{abstract}
A caged compound CeOs$_2$Al$_{10}$, crystallizing in the orthorhombic YbFe$_2$Al$_{10}$-type structure, undergoes a mysterious phase transition at $T_0=29$ K.
We report the results of electron diffraction, magnetization, and magnetoresistance for single crystals.
Superlattice reflections characterized by a wave vector $\bm{q} = (0, -2/3, 2/3)$ observed at 15 K indicate a structural modification in the ordered state.
Activation-type behavior of the electrical resistivity along the three principal axes below 50 K suggests gap opening in the conduction band.
The magnetic susceptibility $\chi = M/B$ is highly anisotropic, $\chi_a>\chi_c>\chi_b$, all of which sharply decrease on cooling below $T_0$.
Furthermore, a metamagnetic anomaly in the magnetization and a step in the magnetoresistance occur at $B=6$-8 T only when the magnetic field is applied parallel to the orthorhombic $c$ axis.
However, $T_0$ hardly changes under magnetic fields up to 14 T, irrespective of the field direction.
By using these data, we present a \textit{B-T} phase diagram and discuss several scenarios for the mysterious transition.
\end{abstract}

\maketitle


\section{Introduction}
Phase transitions in f electron systems are induced by charge, magnetic, and orbital degrees of freedom.
In fact, spin density wave (SDW), charge density wave (CDW), reduced magnetic moment, and unconventional superconducting orderings occur in Ce-, Yb-, and U-based compounds.\cite{Flouquet}
The hybridization of the localized $f$ electrons with conduction electrons (\textit{c-f} hybridization) plays an important role in these transitions as well as a heavy-fermion and intermediate-valence behaviors.
The so-called hidden order in the heavy-fermion superconductor URu$_2$Si$_2$ has attracted much attention.\cite{Shah,Palstra,Maple}
A $\lambda$-type anomaly in the specific heat\cite{Palstra,Maple} indicated a second order transition occurring at $T_0=17.5$ K.
The BCS-type specific heat jump suggested an occurrence of SDW or CDW transition.
The temperature dependence of specific heat below $T_0$ revealed the formation of an energy gap ($\Delta\sim 120$ K) over a part of Fermi surface.
A striking feature is the very small magnetic moment (0.03 $\mu_{\rm B}$/U) observed by neutron scattering experiments in the ordered state\cite{Broholm} which is inconsistent with a significant amount of magnetic entropy of $0.2R\ln 2$ at $T_0$.
Various models have been proposed to explain the hidden order parameter for the phase transition in URu$_2$Si$_2$, but it remains undefined yet.\cite{Harima1}

\begin{figure}[b]
\includegraphics[width=8cm]{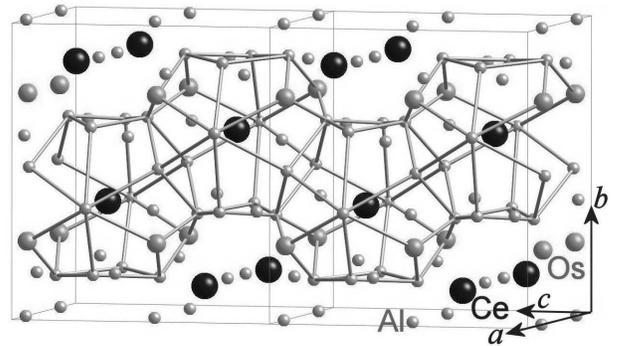}
\caption{\label{structure} Crystal structure of CeOs$_2$Al$_{10}$. Large, middle, and small circles denote Ce, Os, and Al atoms, respectively. The Ce atoms form a zigzag chain along the $c$ axis.}
\end{figure}

Recently, several research groups reported unusual properties of CeT$_2$Al$_{10}$ (T=Fe, Ru, and Os),\cite{MuroFe,Strydom,MuroRu,Nishioka} whose phase transitions for T = Ru and Os resemble that in URu$_2$Si$_2$.
These compounds crystallize in the orthorhombic YbFe$_2$Al$_{10}$-type structure (space group $Cmcm$, No. 63).\cite{Thiede,Tursina}
As shown in Fig. \ref{structure}, the Ce atom is surrounded by 4 T and 16 Al atoms which form a polyhedron.
The nearest-neighbor Ce atoms ($d_{\rm Ce-Ce} = 5.25$ \AA\ for T = Os) construct a zigzag chain along the orthorhombic $c$ axis.
For T = Fe, a broad maximum at 70 K in the magnetic susceptibility is a characteristic of a valence fluctuation compound with the Kondo temperature $T_{\rm K} = 360$ K.\cite{MuroFe}
The thermal activation-type behaviors in both the electrical resistivity and Hall coefficient below 20 K was attributed to the formation of a hybridization gap.
The gap formation in CeFe$_2$Al$_{10}$ has been confirmed by the strong decrease of the $^{27}$Al NMR Knight shift on cooling below 70 K.\cite{Chen}

On the other hand, CeRu$_2$Al$_{10}$ undergoes a phase transition at $T_0=27$ K as indicated by the $\lambda$-type anomaly in the specific heat.\cite{Strydom,Nishioka}
The magnetic entropy up to $T_0$ was estimated to be $0.75R\ln 2$.
By the transition, the Sommerfeld coefficient $\gamma$ is decreased from 246 to 24.5 mJ/K$^2$mol.\cite{Nishioka}
The electrical resistivity shows a thermal activation-type behavior below 40 K, then displays a rapid increase below $T_0$ and a peak at 23 K.
On further cooling, the resistivity decreases gradually.
Similar anomalies have been observed for single-crystal samples.\cite{Nishioka}
The magnetic susceptibility is highly anisotropic $\chi_a>\chi_c>\chi_b$, and $\chi_a(T)$ follows the Curie-Weiss law expected for the trivalent Ce ions.
On cooling below $T_0$, $\chi_a(T)$ suddenly decreases to a constant value.
The $^{27}$Al-NQR signals for Al(1)-(4) sites split into two signals, respectively, whereas no broadening was observed below $T_0$.\cite{Matsumura1}
The gap formation over a portion of Fermi surface is deduced from the temperature dependence of spin-lattice relaxation rate.
These results indicate that the transition is not an antiferromagnetic or SDW one, but is a structural transition which breaks the twofold symmetry along the $a$ axis.
If no structural transition occurred, CeRu$_2$Al$_{10}$ should be a good metal according to a band structure calculation assuming the space group $Cmcm$.\cite{Harima2}

CeOs$_2$Al$_{10}$ also exhibits a phase transition at $T_0=29$ K,\cite{Nishioka} being very close to $T_0$ in CeRu$_2$Al$_{10}$.
However, above $T_0$, the susceptibility of CeOs$_2$Al$_{10}$ shows a pronounced maximum at 45 K which indicates stronger \textit{c-f} hybridization than in CeRu$_2$Al$_{10}$.
Another difference between the two systems appears in the resistivity in the ordered state; the resistivity for T = Os displays a thermal activation-type temperature dependence below 15 K while that for T = Ru shows "metallic" behavior.
Therefore, systematic investigations of CeT$_2$Al$_{10}$ (T = Fe, Ru, and Os) with different values of $T_{\rm K}$ is necessary to reveal the role of the \textit{c-f} hybridization in the mysterious phase transition and gap formation.
In this paper, we report the results of structural and magnetic study via electron diffraction experiments and the measurements of specific heat, magnetization, and magnetoresistance on single crystals of CeOs$_2$Al$_{10}$.

\section{Experimental}
Single crystals of CeOs$_2$Al$_{10}$ were grown using an Al self-flux method in alumina crucible which was sealed in quartz tube under an Ar atmosphere of 1/3 atm.
The crucible was heated to 1100 $^{\circ}$C for 5 h and then cooled slowly to 750 $^{\circ}$C at which point the Al flux was spun off in a centrifuge.
The obtained polyhedral single crystals have typical dimensions of $4\times 3\times 4$ mm$^3$.
Electron-probe microanalyses indicated the composition of Ce : Os : Al = 1 : 1.9(5) : 10.0(5).
Powder x-ray patterns confirmed the orthorhombic YbFe$_2$Al$_{10}$-type structure with lattice parameters $a=9.1386$, $b=10.2662$, and $c=9.1852$ \AA .
It should be noted that the previous report took the opposite notation for $a$ and $c$ axes.\cite{Nishioka}

Transmission electron microscopy observations were performed at the acceleration voltage of 200 kV using JEM2010F equipped with a liquid He cooling stage.
Magnetization $M(B)$ measurements up to 9.5 T were carried out by a capacitive Faraday technique with a field gradient of 10 T/m in the temperature range from 0.3 to 30 K.
The absolute value of $M(B)$ was calibrated using the data obtained by a SQUID magnetometer (Quantum Design MPMS).
The electrical resistivity $\rho(T)$ and magnetoresistance $\rho(B)$ up to 14 T was measured in a longitudinal configuration by an ac four-probe method.
The specific heat $C(T)$ under magnetic fields up to 14 T was measured by a relaxation method.
Measurements of $\rho(B)$ and $C(T)$ were performed using a Quantum Design PPMS.

\section{Experimental Results}
\begin{figure}[t]
\includegraphics[width=8cm]{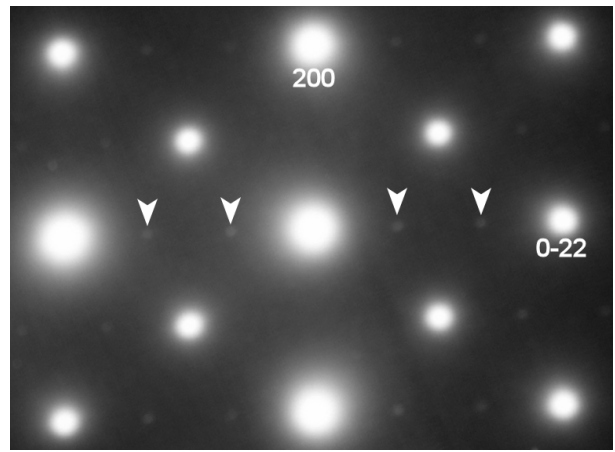}
\caption{\label{EDphoto} [011] zone-axis electron diffraction pattern of CeOs$_2$Al$_{10}$ obtained at 15 K. Superlattice reflections along the [0$-$11] direction are indicated by arrows.}
\end{figure}
\subsection{Electron diffraction}
Figure \ref{EDphoto} shows an selected area electron diffraction pattern of CeOs$_2$Al$_{10}$ at 15 K along [011] zone-axis.
The main reflections are indexed with the orthorhombic $Cmcm$ structure.
In addition to the fundamental reflections, there exist weak superlattice reflections which are characterized by a wave vector $\bm{q} = (0, -2/3, 2/3)$.
This result suggests that the structure at temperatures below $T_0$ is modulated with a threefold periodicity along the [0\=11] direction.

The structural distortion in CeRu$_2$Al$_{10}$ was found by the observation of splitting of the $^{27}$Al-NQR spectra.
The breakdown of the twofold symmetry along the $a$ axis indicated lowering of the space group from $Cmcm$ to either $Cm2m (Amm2)$ or $Pmnm (Pmmn)$.\cite{Matsumura1}
In CeOs$_2$Al$_{10}$, on the other hand, the modulation wave vector $\bm{q}$ maintains the twofold symmetry along the $a$ axis as well as the mirror symmetry perpendicular to the $a$ axis because the $\bm{q}$ vector is normal to the $a$ axis.
However, the $\bm{q}$ vector breaks both the twofold and mirror symmetry for the $b$ and $c$ axis.
This means that the crystal symmetry of CeOs$_2$Al$_{10}$ is distorted to the monoclinic (possibly $C2/m$) structure by the phase transition.
Preliminary x-ray diffraction experiments at low temperatures failed to detect any superlattice diffraction peak.
To determine the crystal symmetry in the ordered state, not only detailed electron diffraction experiments but also neutron diffraction study are necessary.

\subsection{Magnetic susceptibility and resistivity}
\begin{figure}
\includegraphics[width=8cm]{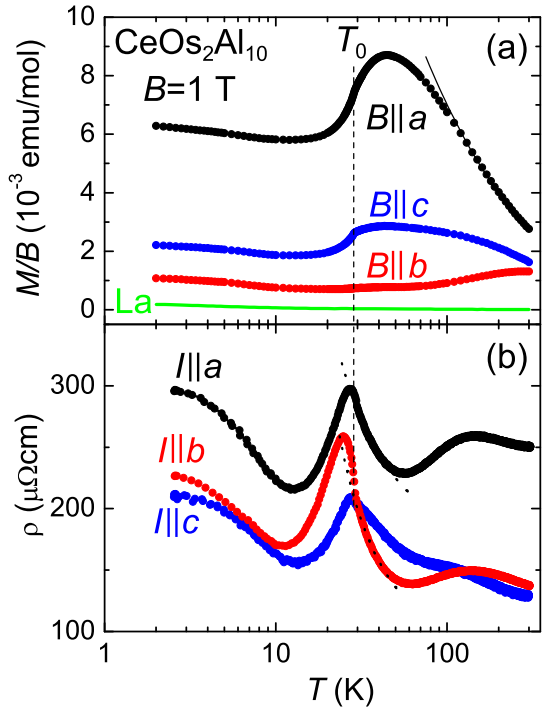}
\caption{\label{suscept} (Color online) (a) Temperature dependence of magnetic susceptibility $M/B$ for CeOs$_2$Al$_{10}$ in magnetic field $B$ applied along the orthorhombic principal axes. The thin solid curve represents the Curie-Weiss fit. The dashed line indicates the transition temperature $T_0$. The susceptibility of LaOs$_2$Al$_{10}$ is also shown by the solid line. (b) Temperature dependence of electrical resistivity of CeOs$_2$Al$_{10}$ for the three current directions. The dotted lines represent the thermally activating behavior for $I||a$ and $I||c$.}
\end{figure}

Figure \ref{suscept}(a) shows the magnetic susceptibility $M/B$($T$) of CeOs$_2$Al$_{10}$ single crystals and LaOs$_2$Al$_{10}$ polycrystal.
The nearly $T$ independent behavior and very small magnitude for LaOs$_2$Al$_{10}$ indicate that the contribution of Os 5$d$ electrons is negligible on the magnetic property of CeOs$_2$Al$_{10}$.
In the whole measured $T$ range, the relation $M_a > M_c > M_b$ holds, that is, the $a$ axis is the easy magnetization axis.
Between 300 and 100 K, $M_a/B(T)$ obeys the Curie-Weiss law, with the effective magnetic moment $\mu_{eff}$ of 2.7 $\mu_B$/Ce and paramagnetic Curie temperature $\theta_{\rm P}$ of $-30$ K.
These values are consistent with those reported.\cite{Nishioka}
A broad peak is depicted in $M_a/B$ at 45 K.
At the transition temperature $T_0=29$ K, $M/B$'s for all axes sharply decrease.
The nearly $T$ independent behavior below 20 K is unlike the continuous decrease in $\chi(T)$ along the easy direction of an antiferromagnet.

The electrical resistivity is also anisotropic as shown in Fig. \ref{suscept}(b).
At high temperatures the $-\ln T$ dependence with a maximum at 140 K is a characteristic of a Kondo-lattice compound.
Both $\rho_a$ and $\rho_b$ exhibit a thermally activation-type behavior $\rho = \rho_0\exp(\Delta/2k_{\rm B}T)$ as shown by the dotted line in Fig. \ref{suscept}(b).
The values of $\Delta/k_{\rm B}$ are obtained as 30 and 50 K for $a$ and $b$ axes, respectively.
This behavior indicates the formation of a pseudo-gap at the Fermi level $E_{\rm F}$ as in CeFe$_2$Al$_{10}$.\cite{MuroFe}
Below $T_0$, $\rho_i$ shows further upturn and then a peak.
Another activation behavior appears on cooling below 14 K, in contrast to the metallic behavior in CeRu$_2$Al$_{10}$.
The semiconducting ground state in CeOs$_2$Al$_{10}$ is consistent with the low value of the Sommerfeld coefficient of 14 mJ/K$^2$mol, which is approximately half of that of CeRu$_2$Al$_{10}$.\cite{Nishioka}

\subsection{Isothermal magnetization}
\begin{figure}
\includegraphics[width=8cm]{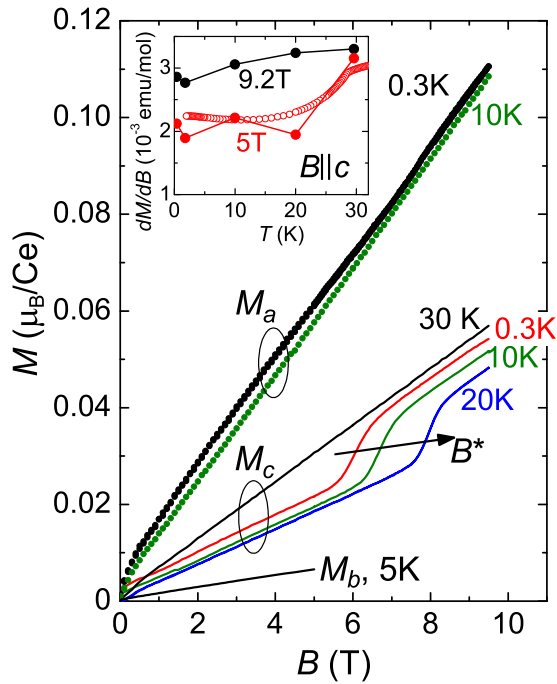}
\caption{\label{magnetization} (Color online) Isothermal magnetization $M_a$, $M_b$, and $M_c$ of CeOs$_2$Al$_{10}$ for $B||a$, $B||b$, and $B||c$, respectively. The temperature variation of the metamagnetic anomaly for $M_c$ is indicated by the arrow. The inset shows the temperature dependence of $dM/dB$ at $B||c = 5$ and 9.2 T. The result of $M/B$ at 5 T measured by a SQUID magnetometer is also shown by open circles.}
\end{figure}

Figure \ref{magnetization} shows the result of $M_i(B)$ measurement for $B$ parallel to the orthorhombic principal axes at various constant temperatures.
Although $M_a$ increases linearly with increasing $B$ up to 9.5 T, $M_c$ exhibits a metamagnetic anomaly near 6 T at 0.3 K.
The critical field $B^*$, which was determined as the peak position in the plot of $dM/dB$ versus $B$, increases from 6.1 T at 0.3 K to 8.0 T at 20 K.
The metamagnetic anomaly disappears at 30 K$>T_0$.
This temperature dependence suggests that the metamagnetic anomaly is not due to the crossing of Ce 4$f$ levels in the orthorhombic crystal field (CF).
In fact, no CF excitations were observed in the energy range up to 20 meV by recent inelastic neutron scattering experiment.\cite{Adroja}
It should be noted that the slope of $M_a$ for $B > 7$ T at 0.3 K is equal to that at 30 K.
This indicates that $dM/dB$ for $B > B^*$ is almost independent of temperature.
This behavior becomes obvious in the $dM/dB$ versus $T$ plot in the inset of Fig. \ref{magnetization}.
At 5 T, $dM_c/dB$ decreases sharply on cooling below $T_0$, in contrast with the very weak decrease in $dM_c/dB(T)$ at 9.2 T.
This contrasting feature in $M_c$ above and below $B^*$ suggests the recovery of the magnetic moment to the value in the normal state by the metamagnetic transition.

\subsection{Longitudinal magnetoresistance}
\begin{figure}
\includegraphics[width=8cm]{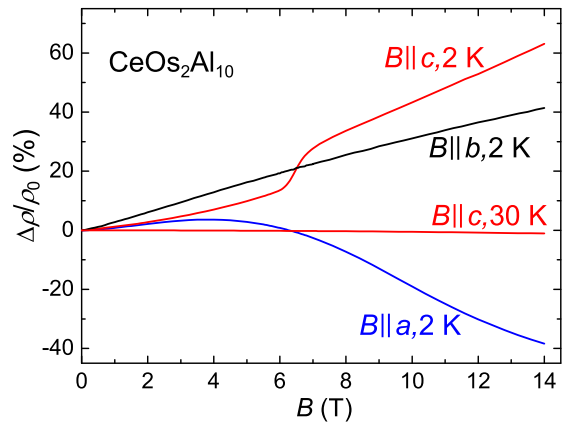}
\caption{\label{magneto} (Color online) Longitudinal magnetoresistance $[\rho(B)-\rho(0)]/\rho(0)$ of CeOs$_2$Al$_{10}$ for $B||a$, $B||b$, and $B||c$ at 2 and 30 K.}
\end{figure}

Figure \ref{magneto} shows the longitudinal magnetoresistance $\Delta\rho/\rho_0 = [\rho(B)-\rho(0)]/\rho(0)$ for $B||a$, $B||b$, and $B||c$ at 2 K.
For $B||a$, easy magnetization axis, $\Delta\rho/\rho_0$ exhibits a broad peak at 4 T and then changes to negative at 6 T.
This behavior is similar to that observed in hybridization-gap systems CeNiSn\cite{Takaba} and CeRhSb,\cite{Yoshino} where the gap is suppressed by the magnetic field applied along the easy magnetization axis.
This similarity corroborates the argument for the pseudo-gap formation over a part of the Fermi surface in CeOs$_2$Al$_{10}$.
Nevertheless, $\Delta\rho/\rho_0$ at 2 K is positive for $B||b$ and $B||c$, as expected for normal metals.
The most important feature in $\Delta\rho/\rho_0$ is the step-like anomaly for $B||c$ at 6.4 T where $M(B||c)$ has the metamagnetic anomaly.
The step in $\Delta\rho/\rho_0$ disappears at 30 K, that is consistent with the disappearance of metamagnetic anomaly at $T>T_0$.

\subsection{Specific heat and \textit{B-T} phase diagram}
\begin{figure}
\includegraphics[width=8cm]{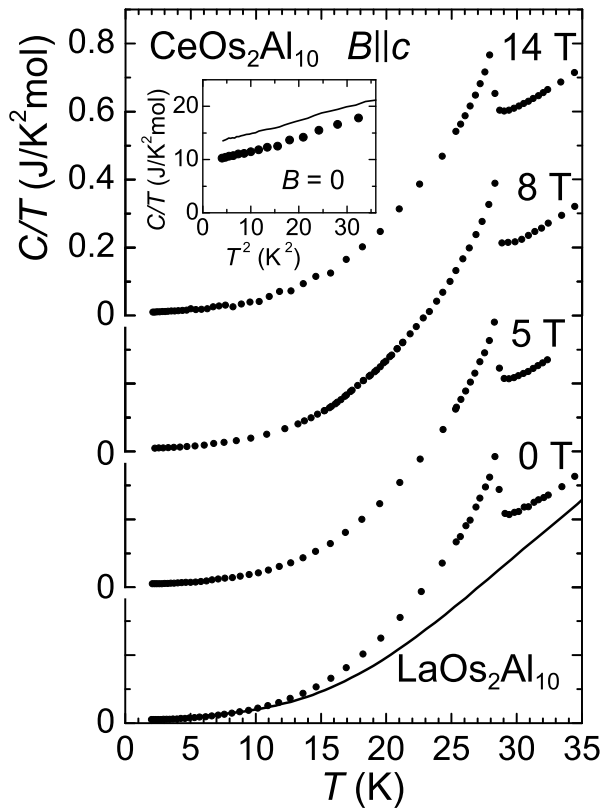}
\caption{\label{Cp} Specific heat divided by temperature $C/T$ for CeOs$_2$Al$_{10}$ measured in various magnetic fields $B = 0$, 5, 8, and 14 T applied along the $c$ axis. A solid line represents the data for LaOs$_2$Al$_{10}$. The inset shows the $C/T$ versus $T^2$ plot at $B = 0$.}
\end{figure}

In order to examine the anisotropic dependence of $T_0$ on magnetic field, we measured $C(T)$ under various constant fields for $B||a$ and $B||c$.
For $B||c$, the result of $C/T$ of CeOs$_2$Al$_{10}$ is plotted versus $T$ in Fig. \ref{Cp}.
As for a reference, the data of the nonmagnetic counterpart LaOs$_2$Al$_{10}$ is shown by the solid line, which has no anomaly above 2 K.
From the $C/T$ versus $T^2$ plot shown in the inset of Fig. \ref{Cp}, we obtained the values of Sommerfeld coefficient $\gamma$ as 10 and 12 mJ/K$^2$mol for CeOs$_2$Al$_{10}$ and LaOs$_2$Al$_{10}$, respectively.
The smaller value of $\gamma$ for CeOs$_2$Al$_{10}$ relative to that of La counterpart supports the existence of pseudo-gap at $E_{\rm F}$.
At 29 K, CeOs$_2$Al$_{10}$ displays a sharp $\lambda$-type anomaly indicative of second-order phase transition. The transition temperature determined by the midpoint of jump in $C(T)$ is $T_0=28.6$ K at $B=0$.
The magnitude of magnetic entropy $S_m(T_0)$ was calculated to be $0.3R\ln 2$ by the integration of $[C_{\rm CeOs_2Al_{10}}-C_{\rm LaOs_2Al_{10}}]/T$, whose value agrees with the reported one.\cite{Nishioka}
With increasing $B$, $T_0$ decreases slowly to 28.2 K at 14 T.
The robustness of the transition against magnetic field suggests that the phase transition in CeOs$_2$Al$_{10}$ is not an antiferromagnetic order.

\begin{figure}
\includegraphics[width=8cm]{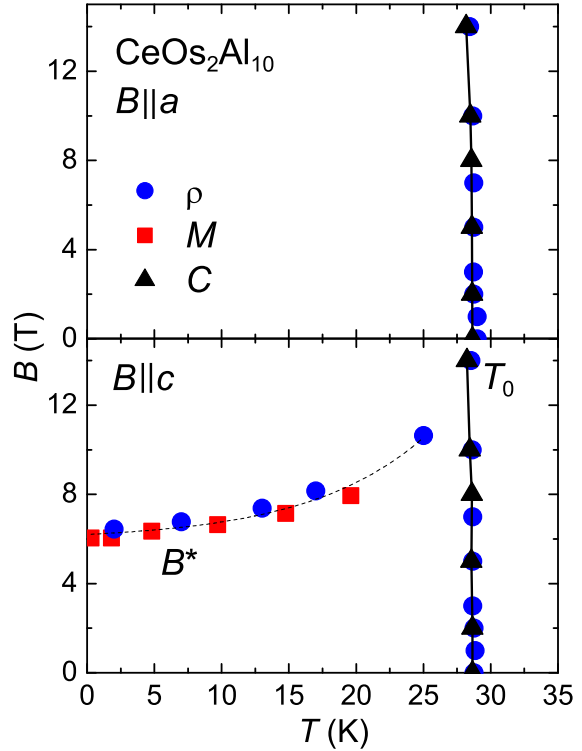}
\caption{\label{PD} (Color online) Magnetic phase diagram of CeOs$_2$Al$_{10}$ for $B||a$ and $B||c$ constructed from the measurements of magnetization, resistivity, and specific heat.}
\end{figure}

  From the results of $C(T, B)$ and $\rho(T, B)$, we constructed magnetic phase diagrams for $B||a$ and $B||c$ as shown in Fig. \ref{PD}.
The magnetic field dependence of $T_0$ for $B||a$ coincides with that for $B||c$.
Thus, $T_0$ hardly changes under magnetic field up to 14 T, irrespective of the field direction.
Measurements of $C(T, B)$ and $\rho(T, B)$ at higher fields are necessary to determine the critical field where the line of $T_0(B)$ terminates.

\section{Discussion}
The absence of phase transition in LaOs$_2$Al$_{10}$ indicates the significant role of the Ce $4f$ electrons in the phase transition of CeOs$_2$Al$_{10}$.
However, the magnetic entropy $S_m(T_0)$ is only 30\% of $R\ln 2$ that is expected for the doublet ground state of localized Ce $4f$ state under the orthorhombic CF.
In the ordered state of CeOs$_2$Al$_{10}$, we observed the superlattice reflections at $\bm{q} = (0, -2/3, 2/3)$ in the transmission electron diffraction patterns.
As for such a transition accompanying a structure modulation, one can enumerate quadrupole, chage/spin Peierls, and CDW transitions.
Conventional magnetic transitions are at variance with the absence of internal field at Al sites.
The quadrupole order of Ce $4f$ electrons is obviously ruled out because the doublet ground state possesses no quadrupole moment in the orthorhombic crystal structure.
For the Peierls transition, one-dimensional instability associated with the dimerization is the prerequisite.
Indeed, the Ce atoms form a zigzag chain along the $c$ axis, and the possibility of spin Peierls transition in CeRu$_2$Al$_{10}$ induced by the dimerization of Ce moments along the zigzag chain has been proposed in recent articles.\cite{Tanida,Hanzawa}
However, the dimerization, if it occurred on the Ce chain, is incompatible with the three-fold periodicity along [0\=11] direction observed as superlattice spots in CeOs$_2$Al$_{10}$.
On the other hand, a CDW transition may be possible in analogy with a Kondo semiconductor CeRhAs.
CeRhAs has been considered as a system exhibiting CDW transitions originating from the fluctuating $f$ electrons.\cite{Sasakawa}
This compound with the orthorhombic $\varepsilon$-TiNiSi-type structure at 300 K is an intermediate-valence compound with the characteristic Kondo temperature $T_{\rm K} = 1200$ K.\cite{Sasakawa}
It undergoes successive phase transitions at $T_1 = 370$ K, $T_2 = 235$ K, and $T_3 = 165$ K.
The superlattice modulations with the wave vectors $\bm{q}_1 = (0, 1/2, 1/2)$, $\bm{q}_2 = (0, 1/3, 1/3)$, and $\bm{q}_3 = (1/3, 0, 0)$ are observed at $T_1$, $T_2$, and $T_3$, respectively, and the gap size is enlarged by the transition at $T_3$.\cite{Matsumura2}
It was pointed out that the transition at $T_3$ is CDW one originating from the charge modulation of Ce ions.
In CeRhAs, there is a zigzag chain of Ce ions along the orthorhombic $a$ axis.
This chain is expected to play an important role in the transition because of the large lattice contraction in the $a$-direction relative to the other directions.
In order to clarify the role of the Ce valence instability and the Ce zigzag chain for the transition in CeOs$_2$Al$_{10}$, a high resolution electron diffraction and neutron scattering study on single crystals are needed.

\section{Summary}
The electron diffraction experiments and the measurements of specific heat, magnetization, and magnetoresistance were performed to study the novel phase transition at $T_0=28.6$ K in CeOs$_2$Al$_{10}$ using single crystals.
The electron diffraction at 15 K revealed the structural modification in the ordered state.
The modulation wave vector indicates a three-fold periodicity along [0\=11] direction.
On cooling below 50 K, a transport gap of 30-50 K was suggested by the activation-type behavior in the electrical resistivity along the three principal axes.
Further development of the pseudo-gap in the density of states associated with the phase transition was confirmed by the sharp decrease in the magnetic susceptibility along the three axes.
The magnetization and magnetoresistance display a metamagnetic anomaly and a step-like increase, respectively, only when the field was applied along the $c$ axis at temperatures below $T_0$.
The negative magnetoresistance in the longitudinal configuration for $B||a > 6$ T indicates that the quasi-gap at the Fermi energy is suppressed by the application of magnetic field along the easy-magnetization axis.
The three-fold periodicity as well as the sharp rise in the resistivity anomaly below $T_0$ favors a CDW for the phase transition.
On the other hand, the very weak decrease in $T_0$ in magnetic fields up to 14 T for both $B||a$ and $B||c$ confirms that the phase transition is not a conventional magnetic order.

\begin{acknowledgments}
We would like to thank J. Yamaura, Z. Hiroi, E. Nishibori, and H. Sawa for x-ray diffraction experiments.
This work was supported by a Grant-in-Aid for Scientific Research on Innovative Areas "Heavy Electrons" (20102004) of the Ministry of Education, Culture, Sports, Science, and Technology, Japan.
\end{acknowledgments}


\begin{thebibliography}{} %
\bibitem{Flouquet} J. Flouquet, in \textit{Progress in Low Temperature Physics}, edited by W. P. Halperin (Elsevier, Amsterdam, 2005), vol. 15, p. 139.
\bibitem{Shah} N. Shah, P. Chandra, P. Coleman, and J. A. Mydosh, Phys. Rev. B \textbf{61}, 564 (2000).
\bibitem{Palstra} T. T. M. Palstra, A. A. Menovsky, J. van den Berg, A. J. Dirkmaat, P. H. Kes, G. J. Nieuwenhuys, and J. A. Mydosh, Phys. Rev. Lett. \textbf{55}, 2727 (1985).
\bibitem{Maple} M. B. Maple, J. W. Chen, Y. Dalichaouch, T. Kohara, C. Rossel, M. S. Torikachvili, M. W. McElfresh, and J. D. Thompson, Phys. Rev. Lett. \textbf{56}, 185 (1986).
\bibitem{Broholm} C. Broholm, J. K. Kjems, W. J. L. Buyers, P. Matthews, T. T. M. Palstra, A. A. Menovsky, and J. A. Mydosh, Phys. Rev. Lett. \textbf{58}, 1467 (1987).
\bibitem{Harima1} H. Harima, K. Miyake, and J. Flouquet, J. Phys. Soc. Jpn. \textbf{79}, 033705 (2010).
\bibitem{MuroFe} Y. Muro, K. Motoya, Y. Saiga, and T. Takabatake, J. Phys. Soc. Jpn. \textbf{78}, 083707 (2009).
\bibitem{Strydom} A. M. Strydom, Physica B \textbf{404}, 2981 (2009).
\bibitem{MuroRu} Y. Muro, K. Motoya, Y. Saiga, and T. Takabatake, J. Phys. Conf. Ser. \textbf{200}, 012136 (2010).
\bibitem{Nishioka} T. Nishioka, Y. Kawamura, T. Takesaka, R. Kobayashi, H. Kato, M. Matsumura, K. Kodama, K, Matsubayashi, and Y. Uwatoko, J. Phys. Soc. Jpn. \textbf{78}, 123705 (2009).
\bibitem{Thiede} V. M. T. Thiede, T. Ebel, and W. Jeitschko, J. Mater. Chem. \textbf{8}, 125 (1998).
\bibitem{Tursina} A. I. Tursina, S. N. Nesterenko, E. V. Murashova, H. V. Chernyshev, H. No\"el, and Y. D. Seropegin, Acta Cryst. E \textbf{61}, i12 (2005).
\bibitem{Chen} S. C. Chen and C. S. Lue, Phys. Rev. B \textbf{81}, 075113 (2010).
\bibitem{Matsumura1} M. Matsumura, Y. Kawamura, S. Edamoto, T. Takesaka, H. Kato, T. Nishioka, Y. Tokunaga, S. Kanbe, and H. Yasuoka, J. Phys. Soc. Jpn. \textbf{78}, 123713 (2009).
\bibitem{Harima2} H. Harima (private communication).
\bibitem{Adroja} D. T. Adroja, A. D. Hiller, P. P. Deen, A. M. Strydom, Y. Muro, J. Kajino, W. A. Kockelmann, T. Takabatake, V. K. Anand, J. R. Stewart, and J. Taylor (unpublished).
\bibitem{Takaba} T. Takabatake, M. Nagasawa, H. Fujii, G. Kido, M. Nohara, S. Nishigori, T. Suzuki, T. Fujita, R. Helfrich, U. Ahlheim, K. Fraas, C. Geibel, and F. Steglich, Phys. Rev. B \textbf{45}, 5740 (1992).
\bibitem{Yoshino} T. Yoshino, T. Takabatake, M. Sera, M. Hiroi, N. Takamoto, and K. Kindo, J. Phys. Soc. Jpn. \textbf{67}, 2610 (1998).
\bibitem{Tanida} H. Tanida, D. Tanaka, M. Sera, C. Moriyoshi, Y. Kuroiwa, T. Takesaka, T. Nishioka, H. Kato, and M. Matsumura, J. Phys. Soc. Jpn. \textbf{79}, 043708 (2010).
\bibitem{Hanzawa} K. Hanzawa, to be published in J. Phys. Soc. Jpn
\bibitem{Sasakawa} T. Sasakawa, T. Suemitsu, T. Takabatake, Y. Bando, K. Umeo, M. H. Jung, M. Sera, T. Suzuki, T. Fujita, M. Nakazima, K. Iwasa, M. Kohgi, Ch. Paul, St. Berger, and E. Bauer, Phys. Rev. B \textbf{66}, 041103(R) (2002).
\bibitem{Matsumura2} M. Matsumura, T. Sasakawa, T. Takabatake, S. Tsuji, H. Tou, and M. Sera, J. Phys. Soc. Jpn. \textbf{72}, 1030 (2003).

\end{thebibliography}
\end{document}